\documentclass[twocolumn,prl,showpacs,preprintnumbers,amsmath,amssymb,superscriptaddress,nofootinbib]{revtex4-1}
\usepackage{graphicx}  % needed for figures
\usepackage{amssymb}   % for math
\usepackage{amsmath}
\usepackage{textcomp}  % for \textmu
\usepackage[colorlinks,citecolor=blue]{hyperref}
\usepackage{color}
\newcommand{\apjl}{Astrophysical Journal, Letters}
\newcommand{\apjs}{Astrophysical Journal, Supplement}

\begin{document}
%\title{Mass measurements of neutron-rich gallium isotopes and their implication on the first \emph{r}-process abundance peak formed in a neutron star merger}

\title{Mass Measurements of Neutron-Rich Gallium Isotopes Refine Production of Nuclei of the First \emph{r}-Process Abundance Peak in Neutron Star Merger Calculations}

\author{M.P. Reiter}
\email[Corresponding author: ]{mreiter@ed.ac.uk}
\altaffiliation{Present address: Institute for Particle and Nuclear Physics (IPNP), University of Edinburgh, Peter Guthrie Tait Road, EH9 3FD, Edinburgh, United Kingdome}
\affiliation{II. Physikalisches Institut, Justus-Liebig-Universit\"{a}t, 35392 Gie{\ss}en, Germany} 
\affiliation{TRIUMF, 4004 Wesbrook Mall, Vancouver, British Columbia V6T 2A3, Canada}

\author{S. Ayet San Andr\'{e}s}
\email[Part of doctoral thesis at JLU Giessen (2018)]{}
    \affiliation{II. Physikalisches Institut, Justus-Liebig-Universit\"{a}t, 35392 Gie{\ss}en, Germany}    
\affiliation{GSI Helmholtzzentrum f\"{u}r Schwerionenforschung GmbH, Planckstra{\ss}e 1, 64291 Darmstadt, Germany}

\author{S. Nikas}
 \affiliation{GSI Helmholtzzentrum f\"{u}r Schwerionenforschung GmbH, Planckstra{\ss}e 1, 64291 Darmstadt, Germany}
 \affiliation{Institut f{\"u}r Kernphysik (Theoriezentrum), Technische Universit{\"a}t Darmstadt,
  Schlossgartenstra{\ss}e 2, 64289 Darmstadt, Germany}

\author{J. Lippuner}
	\affiliation{CCS-2, Los Alamos National Laboratory, P.O. Box 1663, Los Alamos, NM 87545, USA}
    \affiliation{Center for Theoretical Astrophysics, Los Alamos National Laboratory, P.O. Box 1663, Los Alamos, NM 87545, USA}
    \affiliation{Joint Institute for Nuclear Astrophysics, Center for the Evolution of the Elements}  

\author{C. Andreoiu}
    \affiliation{Department of Chemistry, Simon Fraser University, Burnaby, British Columbia V5A 1S6, Canada}

\author{C. Babcock}
    \affiliation{TRIUMF, 4004 Wesbrook Mall, Vancouver, British Columbia V6T 2A3, Canada}

\author{B.R. Barquest}
    \affiliation{TRIUMF, 4004 Wesbrook Mall, Vancouver, British Columbia V6T 2A3, Canada}

\author{J. Bollig}
    \affiliation{TRIUMF, 4004 Wesbrook Mall, Vancouver, British Columbia V6T 2A3, Canada}
    \affiliation{Ruprecht-Karls-Universit\"{a}t Heidelberg, D-69117 Heidelberg, Germany}

\author{T. Brunner}
    \affiliation{TRIUMF, 4004 Wesbrook Mall, Vancouver, British Columbia V6T 2A3, Canada}
    \affiliation{Physics Department, McGill University, H3A 2T8 Montr\'{e}al, Qu\'{e}bec, Canada}

\author{T. Dickel}
    \affiliation{II. Physikalisches Institut, Justus-Liebig-Universit\"{a}t, 35392 Gie{\ss}en, Germany} 
    \affiliation{GSI Helmholtzzentrum f\"{u}r Schwerionenforschung GmbH, Planckstra{\ss}e 1, 64291 Darmstadt, Germany}

\author{J. Dilling}
    \affiliation{TRIUMF, 4004 Wesbrook Mall, Vancouver, British Columbia V6T 2A3, Canada}
    \affiliation{Department of Physics \& Astronomy, University of British Columbia, Vancouver, British Columbia V6T 1Z1, Canada}
    
\author{I. Dillmann}
	\affiliation{TRIUMF, 4004 Wesbrook Mall, Vancouver, British Columbia V6T 2A3, Canada}
    \affiliation{Department of Physics and Astronomy, University of Victoria, Victoria, British Columbia V8P 5C2, Canada}
 
 \author{E. Dunling}
    \affiliation{TRIUMF, 4004 Wesbrook Mall, Vancouver, British Columbia V6T 2A3, Canada}
    \affiliation{Department of Physics, University of York, York, YO10 5DD, United Kingdom}
    
\author{G. Gwinner}
    \affiliation{Department of Physics \& Astronomy, University of Manitoba, Winnipeg, Manitoba R3T 2N2, Canada}
    
\author{L. Graham}
    \affiliation{TRIUMF, 4004 Wesbrook Mall, Vancouver, British Columbia V6T 2A3, Canada}

\author{C. Hornung}
    \affiliation{II. Physikalisches Institut, Justus-Liebig-Universit\"{a}t, 35392 Gie{\ss}en, Germany}

\author{R. Klawitter}
    \affiliation{TRIUMF, 4004 Wesbrook Mall, Vancouver, British Columbia V6T 2A3, Canada}
    \affiliation{Max-Planck-Institut f\"{u}r Kernphysik, Heidelberg D-69117, Germany}

\author{B. Kootte}
    \affiliation{TRIUMF, 4004 Wesbrook Mall, Vancouver, British Columbia V6T 2A3, Canada}
    \affiliation{Department of Physics \& Astronomy, University of Manitoba, Winnipeg, Manitoba R3T 2N2, Canada}

\author{A.A. Kwiatkowski}
    \affiliation{TRIUMF, 4004 Wesbrook Mall, Vancouver, British Columbia V6T 2A3, Canada}
    \affiliation{Department of Physics and Astronomy, University of Victoria, Victoria, British Columbia V8P 5C2, Canada} 

\author{Y. Lan}
    \affiliation{TRIUMF, 4004 Wesbrook Mall, Vancouver, British Columbia V6T 2A3, Canada}
    \affiliation{Department of Physics \& Astronomy, University of British Columbia, Vancouver, British Columbia V6T 1Z1, Canada}

\author{D. Lascar}
    \affiliation{TRIUMF, 4004 Wesbrook Mall, Vancouver, British Columbia V6T 2A3, Canada}
    \affiliation{Center for Fundamental Physics, Northwestern University, Evanston, IL 60208, USA}

\author{K.G. Leach}
    \affiliation{Department of Physics, Colorado School of Mines, Golden, Colorado, 80401, USA}
		
\author{E. Leistenschneider}
		\affiliation{TRIUMF, 4004 Wesbrook Mall, Vancouver, British Columbia V6T 2A3, Canada}
    \affiliation{Department of Physics \& Astronomy, University of British Columbia, Vancouver, British Columbia V6T 1Z1, Canada}

 \author{G. Mart\'inez-Pinedo}
 \affiliation{GSI Helmholtzzentrum f\"{u}r Schwerionenforschung GmbH, Planckstra{\ss}e 1, 64291 Darmstadt, Germany}
 \affiliation{Institut f{\"u}r Kernphysik (Theoriezentrum), Technische Universit{\"a}t Darmstadt,
  Schlossgartenstra{\ss}e 2, 64289 Darmstadt, Germany}

\author{J.E. McKay}    
    	\affiliation{TRIUMF, 4004 Wesbrook Mall, Vancouver, British Columbia V6T 2A3, Canada}
    \affiliation{Department of Physics and Astronomy, University of Victoria, Victoria, British Columbia V8P 5C2, Canada}

\author{S.F. Paul}
    \affiliation{TRIUMF, 4004 Wesbrook Mall, Vancouver, British Columbia V6T 2A3, Canada}
    \affiliation{Ruprecht-Karls-Universit\"{a}t Heidelberg, D-69117 Heidelberg, Germany}

\author{W.R. Pla\ss}
    \affiliation{II. Physikalisches Institut, Justus-Liebig-Universit\"{a}t, 35392 Gie{\ss}en, Germany}
    \affiliation{GSI Helmholtzzentrum f\"{u}r Schwerionenforschung GmbH, Planckstra{\ss}e 1, 64291 Darmstadt, Germany}

\author{L. Roberts}
\affiliation{National Superconducting Cyclotron Laboratory, Michigan State University, 640 South Shaw Lane, East Lansing, Michigan 48824, USA}

\author{H. Schatz}
\affiliation{Joint Institute for Nuclear Astrophysics, Center for the Evolution of the Elements} 
\affiliation{National Superconducting Cyclotron Laboratory, Michigan State University, 640 South Shaw Lane, East Lansing, Michigan 48824, USA}
\affiliation{Department of Physics and Astronomy,Michigan State University, 567 Wilson Road, East Lansing, Michigan 48824, USA}

\author{C. Scheidenberger}
    \affiliation{II. Physikalisches Institut, Justus-Liebig-Universit\"{a}t, 35392 Gie{\ss}en, Germany}
    \affiliation{GSI Helmholtzzentrum f\"{u}r Schwerionenforschung GmbH, Planckstra{\ss}e 1, 64291 Darmstadt, Germany}

 \author{A. Sieverding}
  \affiliation{GSI Helmholtzzentrum f\"{u}r Schwerionenforschung GmbH, Planckstra{\ss}e 1, 64291 Darmstadt, Germany}
 \affiliation{Institut f{\"u}r Kernphysik (Theoriezentrum), Technische Universit{\"a}t Darmstadt,
  Schlossgartenstra{\ss}e 2, 64289 Darmstadt, Germany}
 \affiliation{School of Physics and Astronomy, University of Minnesota, 116 Church Street SE, Minneapolis 55455, USA}
 
\author{R. Steinbr\"{u}gge}
		\altaffiliation{Present address: Deutsches Elektronen-Synchrotron DESY, Notkestr. 85, 22607 Hamburg, Germany}
    \affiliation{TRIUMF, 4004 Wesbrook Mall, Vancouver, British Columbia V6T 2A3, Canada}
    
\author{R. Thompson}
    \affiliation{Department of Physics and Astronomy, University of Calgary, Calgary, Alberta T2N 1N4, Canada}    

\author{M.E. Wieser}
    \affiliation{Department of Physics and Astronomy, University of Calgary, Calgary, Alberta T2N 1N4, Canada}
    
\author{C. Will}
    \affiliation{II. Physikalisches Institut, Justus-Liebig-Universit\"{a}t, 35392 Gie{\ss}en, Germany}  

\author{D. Welch}
\affiliation{National Superconducting Cyclotron Laboratory, Michigan State University, 640 South Shaw Lane, East Lansing, Michigan 48824, USA}

\date{\today}

\begin{abstract}
We report mass measurements of neutron-rich Ga isotopes $^{80-85}$Ga  with TRIUMF's Ion Trap for Atomic and Nuclear science (TITAN). The measurements
%, employing a multiple-reflection time-of-flight mass spectrometer, 
determine the masses of $^{80-83}$Ga in good agreement with previous measurements. The masses of $^{84}$Ga and $^{85}$Ga were measured for the first time. Uncertainties between $25-48$~keV were reached. The new mass values reduce the nuclear uncertainties associated with the production of A $\approx$ 84 isotopes by the \emph{r}-process for astrophysical conditions that might be consistent with a binary neutron star (BNS) merger producing a blue kilonova. Our nucleosynthesis simulations confirm that BNS merger may contribute to the first abundance peak under moderate neutron-rich conditions with electron fractions $Y_e=0.35-0.38$. %These conditions place the \emph{r}-process path closest to stability and may allow for a nuanced investigation of the formation of the A~$=$~80~and~84 abundance maxima of the first \emph{r}-process abundance peak.

%The formation of the first emph{r}-process peak  electron fractions $Y_e$, which contribute to the formation of the first \emph{r}-process abundance peak, to $Y_e=0.5-0.38$. 

\end{abstract}

\maketitle

\section{Introduction}

Since the first discovery of a binary neutron-star (BNS) system (PSR1916+16)%by Hulse and Taylor 
\cite{hulse1975discovery}, the merger of two neutron stars has been considered a promising site for the production of heavy elements by the rapid neutron capture process, \emph{r}-process \cite{Lattimer.Schramm:1974,lattimer76,Eichler.Livio.ea:1989,rosswog99,freiburghaus:99,Arnould2007}. 
%This process is currently considered to be responsible for the formation of roughly half of the stable elements heavier than iron in the universe \cite{Arnould2007}. %However, an observation of the \emph{r}-process at its sites was elusive until very recently.
%\cite{argast:04, shen:14, voort:15, ramirez-ruiz:15} Covino2017
The \emph{r}-process in BNS mergers provides a unique electromagnetic signature %due to the radioactive decay of freshly synthesized material 
known as kilonova/macronova \cite{lipacz97,metzger2010electromagnetic,Roberts2011,Bauswein:13,Fernandez2016}. Except for a few candidates e.g. \cite{2009ApJ...696.1871P,2013Natur.500..547T,2013ApJ...774L..23B,2015ApJ...811L..22J,2015NatCo...6E7323Y} such signatures were not clearly observed.
The situation changed with the observations of the gravitational waves from the BNS merger (GW170817) \cite{abbott2017gw170817,2041-8205-848-2-L13} 
%on August 17, 2017,
%the associated short gamma-ray burst (GBR170817A) by the Fermi Gamma-Ray Burst Monitor \cite{Goldstein2017,Racusin2017} 
and the subsequent detection of the electromagnetic counterpart (AT2017gfo) \cite{2041-8205-848-2-L12}. The optical, infrared and ultraviolet spectra and their evolution agree well with the macronova/kilonova model, constituting first direct evidence that heavy elements, including the lanthanide region, were synthesized by the \emph{r}-process \cite{Tanvir2017,Chornock2017,Pian2017,Kasen2017}. The early blue emission \cite{evans2017swift,Smartt2017,Troja2018,Nicholl2017}, consistent with electron fractions of around $Y_e\approx 0.25-0.4$~\cite{Rosswog2018,Wanajo2018,Wu2018}, suggests the production of intermediate mass \emph{r}-process nuclides with masses below $A < 140$. However, there has been in general no direct evidence for the production of elements of the first \emph{r}-process abundance peak, except one recent study, which concludes the identification of strontium in a reanalysis of the AT2017gfo specra \cite{watson2019identification}.  %\cite{Smartt2017,Chornock2017} 

%Similar conditions are also expected in magnetorotational supernovae~\cite{Winteler.Kaeppeli.ea:2012,Nishimura2015}, that are also considered as possible sites for the r process.

\emph{R}-process  %alpha particles are not present and 
nucleosynthesis proceeds by $(n,\gamma)$ neutron captures in competition with $(\gamma,n)$ photodissociation reactions and $\beta$-decay. Nuclei around the closed neutron shells serve as waiting points and sensitivity studies have shown strong dependence of the final abundance pattern on nuclear masses \cite{Brett2012,doi:10.1063/1.4867193,Mendoza.Wu.ea:2015,mumpower2015impact2_mass,mumpower2015impact_mass}, $\beta$-decay rates \cite{mumpower2014sensitivitybeta,Marketin.Huther.ea:2016,mumpower2016impact}, $\beta$-delayed neutron emission~\cite{Kratz1993}, fission properties \cite{Eichler.Arcones.ea:2015,Goriely2015,Giuliani.Martinez.ea:2018}, $(n,\gamma)$ reaction rates \cite{mumpower2016impact}% of these isotopes
, as well as to statistical quantities like strength functions and level densities \cite{snikas_in_prep}. 

Simulations show that magnetorotational supernovae~\cite{Winteler.Kaeppeli.ea:2012,Nishimura2015} and BNS mergers~ \cite{lippuner2015r,martin2015neutrino} can produce the first \emph{r}-process peak under moderate entropy, entropy per nucleon $\approx 10\,k_B/\mathrm{nucleon}$, and moderately neutron-rich conditions, electron fractions $Y_e \approx 0.35$.

%In order to investigate 
To cast more light on the formation of the first \emph{r}-process abundance peak and investigate whether the ejecta of a BNS merger can indeed be one of the possible sites for the formation of $A \approx 80 - 84$ \emph{r}-process elements is of general interest. %due to the lack of direct signatures. 
 This requires BNS merger \emph{r}-process simulations with accurate nuclear physics properties. The formation of the first abundance peak offers a unique opportunity for precision studies, because the \emph{r}-process runs closest to stability where a majority of nuclear properties have been experimentally measured. However to understand the synthesis of $A \approx 84$ nuclei in \emph{r}-process models precise masses of neutron-rich Ga, Ni, Cu and Zn isotopes are needed \cite{Brett2012, doi:10.1063/1.4867193}. 
 
 Here we present the first experimental results for the masses of $^{84,85}$Ga. They significantly reduce the nuclear physics uncertainties for the synthesis of $A \approx 84$ nuclei in \emph{r}-process models and allow a systematic investigation of the formation of the first \emph{r}-process peak.

\section{Experimental Description}

Neutron-rich Ga isotopes were produced by a $\approx 480\,\text{MeV}$, $10\,\text{\textmu A}$ proton beam impinging on a UC$_{x}$ target \cite{Kunz2013} at the ISAC facility \cite{Dombsky2000}. %TRIUMF's Ion Guide Laser Ion Source (IG-LIS) \cite{Raeder2014} was used to suppress surface-ionized contaminants and to resonantly laser ionize the Ga isotopes of interest \cite{LI201374}. 
The continuous, mass-separated, beam from TRIUMF's Ion Guide Laser Ion Source \cite{Raeder2014} was accumulated and bunched in TITAN's Radio-Frequency-Quadrupole (RFQ) cooler-buncher \cite{Brunner2012a} and ion bunches were sent to the Multiple-Reflection Time-of-Flight Mass-Spectrometer and isobar separator (MR-TOF-MS)\cite{Jesch2015,PhysRevLett.120.062503,PhysRevC.98.024310},
%, capable of performing mass measurements of low count rate species with relative uncertainties in the $\delta m/m \sim 10^{-7}$ range \cite{PhysRevLett.120.062503,PhysRevC.98.024310}. %The TITAN MR-TOF-MS is based on the system operated at the FRS at GSI \cite{Yavor2015,Dickel2015}.  
operated at a $20\,\text{ms}$ cycle time. The radioactive ion beam (RIB), containing predominantly singly-charged Rb, Br and Ga, was captured in the gas-filled RFQ system of MR-TOF-MS and transported to a dedicated injection trap. %
Natural Rb ions from a thermal ion source were merged with the RIB via an RFQ switch yard \cite{Plass2015} to provide independent calibration ions. 

%	\begin{figure}[t]
%    \begin{center}
%        \includegraphics[width=0.9\columnwidth]{g4114.png}
%        \caption{Overview of the TITAN mass measurement facility, subsystems not used in this experiment are shown in light gray.}
%        \label{fig:titan}
%    \end{center}
%\end{figure}

\begin{figure}[tb]
    \begin{center}
        \includegraphics[width=\columnwidth]{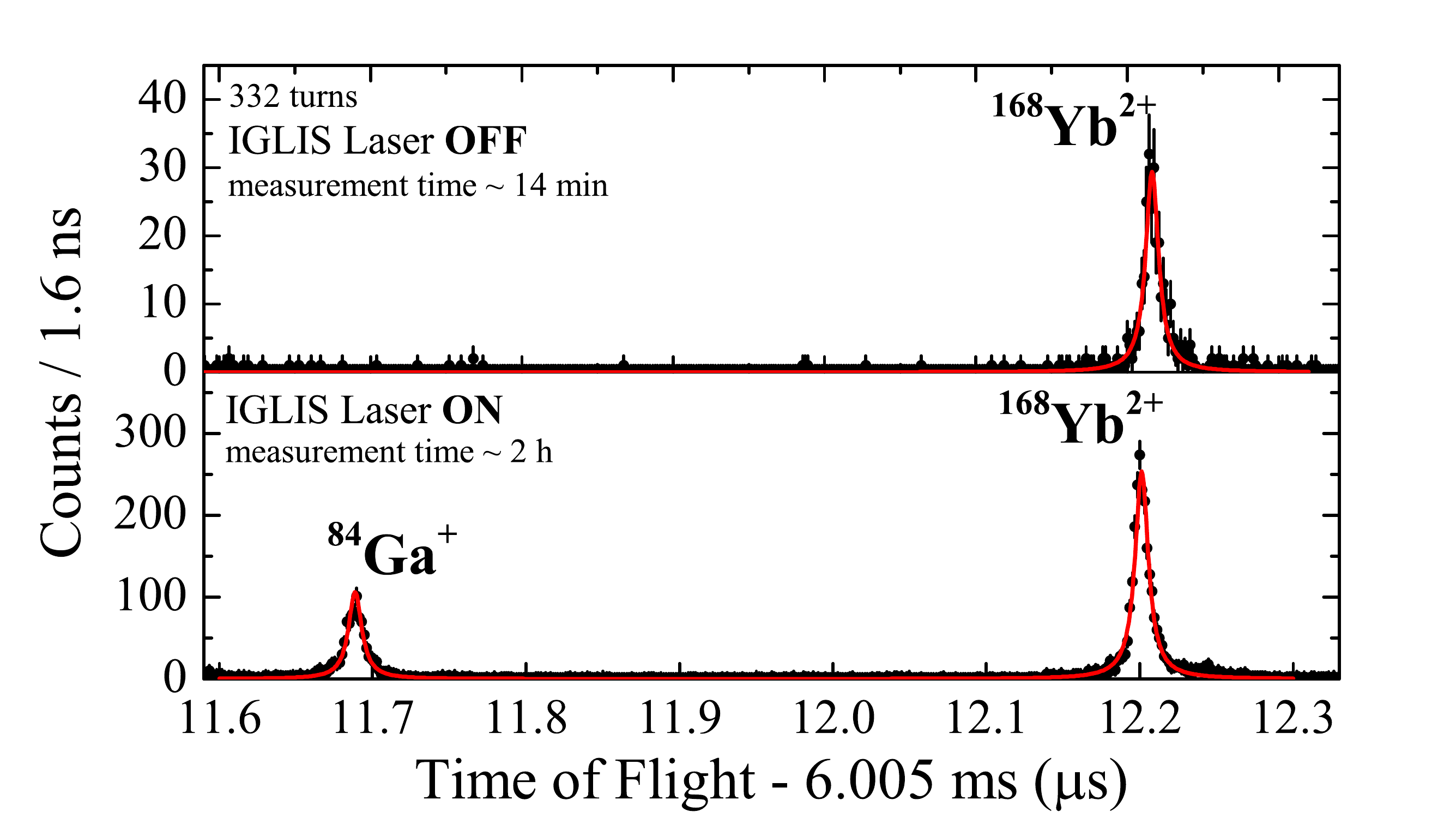}
        \caption{Time-of-flight spectra obtained with the MR-TOF-MS around $^{84}$Ga$^{+}$ confirming the identification of $^{84}$Ga$^{+}$ by blocking the resonant IG-LIS laser. The lower spectra shows $\approx 1200$ $^{84}$Ga$^{+}$ ions from which the mass of $^{84}$Ga has been determined. The red lines are fits to the data using Lorentzian peak shapes.}
        \label{fig:ga84}
    \end{center}
\end{figure}

In order to reach high resolving powers %in the time-of-flight mass spectrometer, 
the flight path is extended by storing ions for multiple reflections between two electrostatic ion mirrors \cite{Wollnik1990}. The spectrometer herein is based on \cite{Yavor2015,Dickel2015}, using a dynamic time-focus-shift \cite{Dickel2017a}.  
The ions were kept between $330$ and $372$ turns in the mass analyzer resulting in flight times between $6.00\,\text{ms}$ and $6.82\,\text{ms}$ and yielding mass resolving powers of $m/\Delta m\approx \text{200\:000}$. %FWHM. %
The number of turns was chosen such that natural Rb ions did not interfere with the RIB ions and arrive outside of the time-of-flight window of the RIB species. %
The ions were detected by a MultiChannelPlate (MCP) detector and the time of flight was recorded using a time-to-digital converter (TDC, ORTEC 9353).

\section{Data Analysis}
%\label{sec:analysis}
In the MR-TOF-MS, the mass of an ion is determined based on the non-relativistic relation between the mass $m$, the charge $q$, and the time of flight $t_\mathrm{tof}$ needed to travel a certain flight path, resulting in 
$
m/q = c (t_\mathrm{tof}-t_{0})^{2}
$.
The measured time of flight $t_\mathrm{tof}$ is the sum of the real time of flight of the ion and a constant delay $t_{0}$ caused by signal propagating times. %through cables and electronic. 
Measuring the time of flight of one or more reference masses allows determination of the calibration parameters. The delay $t_{0}$ depends on the system and data acquisition and can be determined offline from an independent calibration. It was determined from $^{85}$Rb$^{+}$ and $^{87}$Rb$^{+}$ ions %undergoing only one time-focus-shift turn 
and amounted to $t_{0}=116(3)$~ns. For the calibration of $c$ a dominant species from the RIB was chosen; see Tab.~\ref{tab:values}. To account for time-of-flight drifts, resulting from temperature changes and instabilities, a time-dependent calibration was used %\cite{Ebert2016}. 
\cite{PhysRevC.99.064313}.
\newcommand{\0}{\phantom{0}}
\newcommand{\p}{\phantom{.}}
\newcommand{\ph}{\phantom{80m,}}
\newcommand{\m}{\phantom{-}}
\newcommand{\tms}{\phantom{{}\times 10^0}}
\begin{table*}[tb]
  \centering
  \caption{Mass measurements of singly-charged Ga isotopes performed during this experiment using TITAN's MR-TOF-MS in comparison to the values reported in AME2016 \cite{1674-1137-41-3-030003}, the $\#$ indicates extrapolated values therein. In addition, the half-life taken from \cite{audi2017nubase2016}, the number of isochronus turns (IT) the Ga ions were stored in the analyzer and the respective calibration species are given.}
  		\begin{tabular*}{\textwidth}{@{\extracolsep{\fill}} lcclcll}
    \hline
		\hline
    \multicolumn{1}{c}{Species} & $t_{1/2}$ \cite{audi2017nubase2016} & No. of & \multicolumn{1}{c}{Calibrant} & Mass Excess$_{\mathrm{TITAN}}$ & \multicolumn{1}{c}{Mass Excess$_{\mathrm{AME2016}} $} & \multicolumn{1}{c}{Difference}  \\
         & (ms) &   IT$_{\mathrm{Ga}}$ & &  (keV/c$^2$) & \multicolumn{1}{c}{(keV/c$^2$)} & \multicolumn{1}{c}{(keV/c$^2$)}   \\
    \hline
  $^{\ph80}$Ga$^\ast$ & 1900(100)    &366& $^{\ph80}$Ge  & $-$59\,212(48) & $-$59\,223.7(2.9)      & \0$-12(48)$    \\
  $^{\ph81}$Ga         & 1217(5)\p\0    &373& $^{\ph81}$Br  & $-$57\,616(31) & $-$57\,628(3)          & \0$-12(31)$    \\
  $^{\ph82}$Ga$^\ast$  & \0599(2)\p\0   &382& $^{\ph82}$Rb$^\diamond$ & $-$52\,974(31) & $-$52\,930.7(2.4)      & $\m\043(31)$   \\
  $^{\ph83}$Ga         & \0308(10)          &333& $^{\ph83}$Rb  & $-$49\,258(25) & $-$49\,257.1(2.6)      & $\m\0\01(25)$  \\
  $^{\ph84}$Ga         & \0\085(10)        &332& $^{\ph84}$Rb  & $-$44\,094(30) & $-$44\,090(200)$\#$ & $\m\0\04(202)\#$ \\
  $^{\ph85}$Ga         & \0\092(4)  &360& $^{\ph85}$Rb  & $-$39\,744(37) & $-$39\,850(300)$\#$ & $-106(302)\#$    \\
 % $^{\ph85}$Ga         & \0\092\p\0 & $\sim 1 \tms $     &360& $^{\ph85}$Rb  & $-$39,744(32) & $-$39,850(300)$^\dag$ & $-106(302)$    \\
    \hline
    \hline
  \end{tabular*}
  \vspace{0.5ex}\par
  
  ~\makebox[0pt]{$^\ast$} ~These measurements were affected by an unresolved isomeric state in either the Ga isotope of interest or the calibration species, see text for description. \hspace{\fill}
	
	~\makebox[0pt]{$^\diamond$} ~Assuming the measured state in $^{82}$Rb is dominantly the isomer at $69$~keV, as suggested by spectroscopy at ISAC Yield Station, the mass of $^{82}$Ga is $-$52\,939(23) keV/c$^2$.  \hspace{\fill}
  
%  ~\makebox[0pt]{$\#$} ~These are extrapolated values, that were unmeasured prior to our measurements.\hspace{\fill}
 \label{tab:values}%
\end{table*}

The corresponding Ga peaks %at all mass units 
could be clearly identified by their time-of-flight and by performing a measurement with and without the resonant laser ionization step \cite{LI201374}, shown in Fig. \ref{fig:ga84} for $^{84}$Ga. To account for peak shape dependent effects, particular for nearby or overlapping peaks, two independent analyses were performed, using Gaussian and Lorentzian line shapes, similar to \cite{PhysRevC.98.024310}. %The same peak width was used to fit all peaks within one spectra. 
The final error on the mass value was calculated by quadratically adding: (a) the uncertainty from the fitting algorithm (b) the statistical uncertainty of the ion of interest, (c) the uncertainty of the calibration peak and its uncertainty reported in the AME2016 \cite{1674-1137-41-3-030003} and (d) a systematic uncertainty of $\delta m/m_{syst}~=~3\times 10^{-7}$ \cite{Will2017thesis}. The systematic uncertainty was redetermined from accuracy measurements of $^{85}$Rb$^{+}$ and $^{87}$Rb$^{+}$. In order to eliminate possible effects from ion-ion interactions the total number of ions was kept below two detected ions per cycle.

For the $^{80,82}$Ga measurements systematic effects, arising from unresolved isomeric states, had to be taken into account. An isomeric state in $^{80}$Ga at $22.4$~keV \cite{PhysRevC.87.054307} and in the calibration species $^{82}$Rb at $69$~keV \cite{NSR1967VR07} could not be resolved. The final mass values were corrected and an additional uncertainty was added, according to the procedures in AME2016 Appendix B.1 \cite{Huang2017c}. %for the treatment of unresolved isomeric states. 

%\section{Discussion of Experimental Results}
%\begin{figure}[tb]
%    \begin{center}
%        \includegraphics[width=\columnwidth]{Ga_overview3.eps}
%        \caption{Mass excess of neutron-rich gallium isotopes determined in this TITAN experiment in comparison to the literature values reported in \cite{Huang2017c}, shown as gray band. For the nuclides previously measured by JYFLTRAP \cite{Hakala2008} an average agreement well within one $\sigma$ can be seen.}
%        \label{fig:compar}
%    \end{center}
%\end{figure}

\section{Results and Discussion}
The final results are summarized and compared to the values given in the AME2016 \cite{Huang2017c} in Tab.~\ref{tab:values}. %Fig.~\ref{fig:compar} and 
The mass values of $^{80,81,83}$Ga agree well with the AME2016 values, which are based on measurements performed by JYFLTRAP \cite{Hakala2008}. Our result for $^{82}$Ga deviates by $1.3\sigma$ from the previous measurement. Assuming the measured state in the calibration species $^{82}$Rb is dominantly the isomer at an excitation energy of $69$~keV, as suggested based on spectroscopy at ISAC Yield Station, the mass of $^{82}$Ga results in $-$52\,939(23) keV/c$^2$, which is in good agreement with the JYFLTRAP result. 
The masses of $^{84,85}$Ga were measured for the first time and are compared to extrapolations in Tab.~\ref{tab:values}. %given in the AME2016 \cite{1674-1137-41-3-030003}. 
\begin{figure}[tb]
    \begin{center}
        \includegraphics[width=\columnwidth]{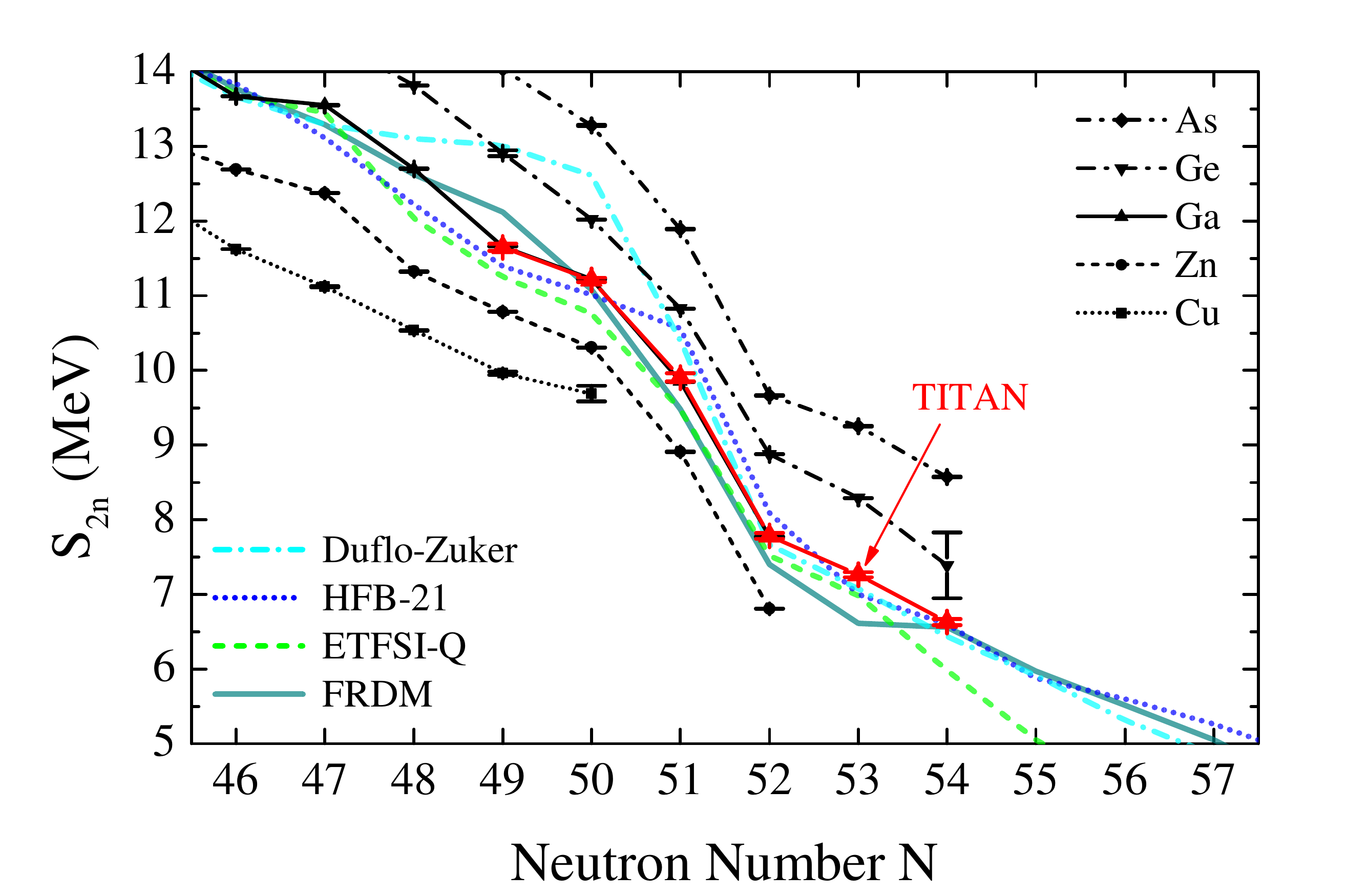}
        \caption{(Color online) Experimental two-neutron separation energies $S_{2\text{n}}$ for $Z=30-33$ (Zn to As) as a function of neutron number taken from AME2016 \cite{1674-1137-41-3-030003}, including \cite{PhysRevLett.119.192502}. For comparison, $S_{2\text{n}}$ values based on the new TITAN masses (red) and based on commonly used mass models (FRDM \cite{MOLLER20161}, Duflo-Zucker \cite{PhysRevC.52.R23}, ETFSI-Q \cite{pearson1996nuclear}, HFB-21 \cite{goriely2013further}) are shown.}
        \label{fig:s2n}
    \end{center}
\end{figure}  

Based on the mass values $M$, we calculate the two-neutron separation energy $S_{2\text{n}}(N,Z) = M(Z,N-2)c^2 + 2 M_{n}c^2 - M(N,Z)c^2$, with $M_{n}$ the mass of the neutron, and compare it in Fig.~\ref{fig:s2n} to the neighboring isotopic chains. The drop in $S_{2\text{n}}$, associated with the closed neutron shell at $N=50$, can be seen in the Ga isotopic chain \cite{Hakala2008}. The new $S_{2\text{n}}$ values for $^{84,85}$Ga confirm the recurrence to a smooth trend beyond the $N=50$ shell closure and bring the Ga isotopic chain in line with the neighboring Ge and As chains. 

We compare the experimental $S_{2\text{n}}$ values to values based on commonly used mass models (FRDM \cite{MOLLER20161}, Duflo-Zucker \cite{PhysRevC.52.R23}, ETFSI-Q \cite{pearson1996nuclear}, HFB-21 \cite{goriely2013further}). For the Ga isotopes in this region FRDM and HFB-21 show overall good agreement, whereas ETFSI-Q systematical predicts less binding and Duflo-Zucker over-predicts the strength of the $N=50$ shell closure. Beyond $N=54$, FRDM, HFB-21 and Duflo-Zucker all predict a continuation of the smooth trend.

\section{Astrophysical Implications}

    To systematically study the formation of $A \approx 84$ nuclei, $(n,\gamma)$ and $(\gamma,n)$ reaction rates
    corresponding to the 
    %extrapolated AME2016 \cite{1674-1137-41-3-030003} 
    mass values of $^{84,85}$Ga were calculated using the %well established 
    %To quantify the impact of the experimental $^{84,85}$Ga mass values on the first \emph{r}-process peak and study the formation of $A \approx 84$ nuclei, $(n,\gamma)$ and $(\gamma,n)$ reaction rates corresponding to the extrapolated AME2016 \cite{1674-1137-41-3-030003} mass values of $^{84,85}$Ga were calculated using the 
		Hauser-Feshbach statistical code TALYS \cite{koning2007talys}. 
    The resulting cross-sections were initially used in two different nuclear reaction network codes, \emph{GSINet} \cite{Mendoza.Wu.ea:2015} and \emph{SkyNet} \cite{lippuner:17}, to calculate the \emph{r}-process abundances.
    Comparing final abundances from the two network codes showed that both predict almost identical results, highlighting the robustness of the network codes themselves.  % for the \emph{r}-process abundance in the region of the first \emph{r}-process abundance peak.
    %Simulations of wind ejecta indicate that less neutron-rich ejecta, which are
    %expected at late times, can contribute substantially to the production of
    %isotopes of the first \emph{r}-process peak
    %\cite{perego2014neutrino,martin2015neutrino} and as such have been considered.

	\begin{figure}[tb]
    \begin{center}
        \includegraphics[width=\columnwidth]{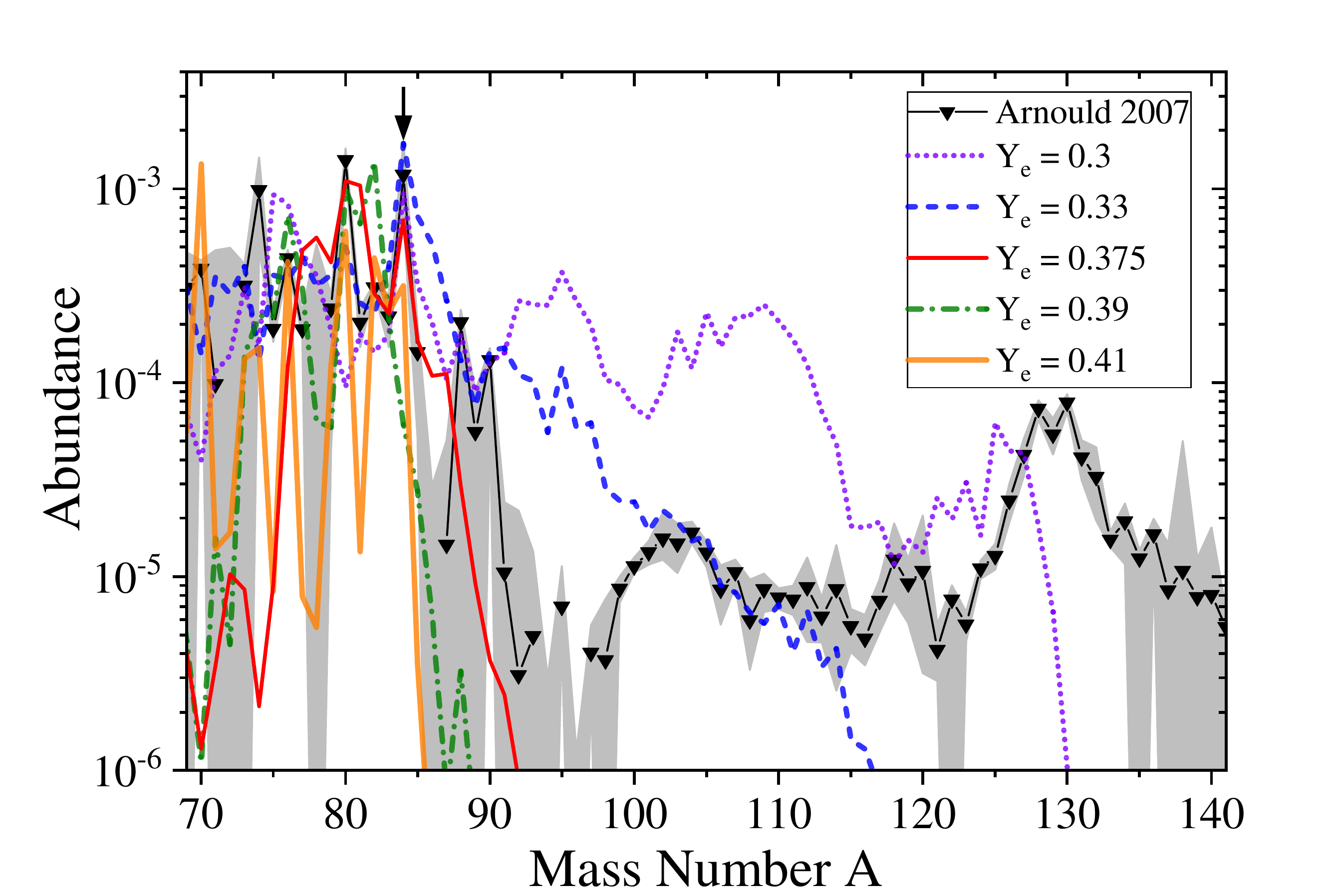}
        \caption{(Color online) Solar \emph{r}-process abundance, with uncertainty shown as gray band, in comparison to the abundance resulting from neutron star merger network calculations for different $Y_e$ using \emph{GSINet}. The individual abundance curves are shown with equal weights. The \emph{r}-process abundance \cite{Arnould2007} has been scaled to match the average production of $^{82}$Se. The arrow indicates the $A=84$ abundance maximum of the first \emph{r}-process abundance peak.}
        \label{fig:solar_r_compare_Ye}
    \end{center}
\end{figure}

 \begin{figure}[tb]
 %   \includegraphics[width=\linewidth]{error_bar_plot.pdf} 
		 %\includegraphics[width=\columnwidth]{3sigma_and_gausian.pdf} 		
	%	\includegraphics[width=\columnwidth]{error_bar_plot_mod_2019-09-30.pdf} 
	%previous zoomed in error bar plot
		\includegraphics[width=\columnwidth]{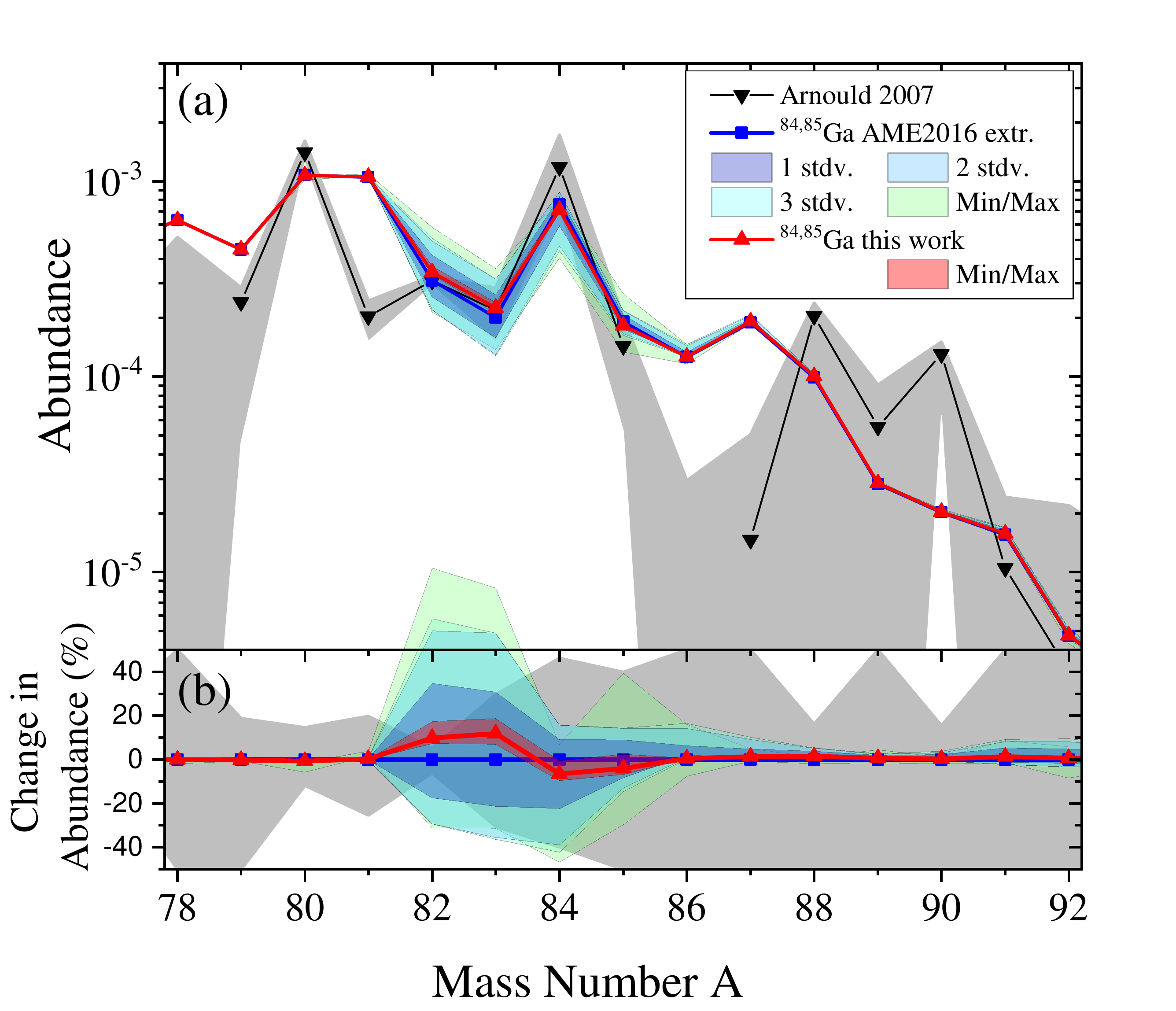} 
	% zoomed out error bar plot 
	%	\includegraphics[width=\columnwidth]{error_bar_plot_mod_2019-06-25eps.eps} 
    \caption{%\todo{update to PRL standard font = Times New Roman} 
    (Color online) (a) Final abundances averaged over calculations with $Y_e=0.35-0.38$
   compared to the solar \emph{r}-process abundance \cite{Arnould2007}, %\cite{Goriely1999}, 
	with uncertainty shown as gray band. The colored bands show the one, two and three sigma change in calculated production, as well as the maximum and minimum abundance from the Monte-Carlo variation of the nuclear masses of $^{84,85}$Ga following a Gaussian distribution with $\sigma$ of $200$ and $300$~keV, respectively.
%Here, the dashed blue line shows that within the uncertainty a relatively flat abundance pattern without a distinct peak at $A=84$ would be allowed, this is ruled out taking the new masses of $^{84,85}$Ga into account (red line with circles). 
 For the new mass values only the maximum and minimum abundance band from the variation within their uncertainty is shown.  
(b) Change, in percent, of the abundance pattern as a result of using the mass values from this work compared to the extrapolations given in the AME2016.    
   }
    \label{fig:nucleo_error_bar}
   \end{figure}
	
		The masses of $^{84,85}$Ga modify the reaction rates around $^{83-86}$Ga; nuclei not affected were taken from JINA REACLIB \cite{cyburt2010jina}. Where available, experimental masses from AME2016 were used, otherwise masses based on the FRDM mass model \cite{MOLLER20161} were taken (with exception of $^{84,85}$Ga). 
		 To quantify the uncertainty of the final abundance associated with the mass values of $^{84,85}$Ga we use a Monte Carlo type approach. The masses of $^{84,85}$Ga were randomly varied within a normal distribution with $\sigma$ according to the uncertainty of their extrapolated mass values given in the AME2016 \cite{1674-1137-41-3-030003}. For a set of one hundred possible combinations of mass values drawn from the uncertainty distribution $(n,\gamma)$ cross-sections were calculated. Combinations that would result in inverted odd-even effects for the one-neutron separation energies in the Ga isotopic chain were excluded. By using each combinatioin in a \emph{GSINet} calculation an estimate for the overall uncertainty of the final abundances was obtained. The procedure was repeated using the new $^{84,85}$Ga mass values and respective uncertainties.  
		%During the estimation of the uncertainty of the production based on the mass values, AME2016 predictions were used for $^{84,85}$Ga. 
		%The values based on FRDM were scaled to the last experimentally known isotope to create a smooth mass surface. 
		$\beta$-decay rates and $\beta$-delayed neutron emission branches were taken from NUBASE \cite{audi2017nubase2016}, including recent measurements \cite{PhysRevC.95.054320,PhysRevC.97.054317,PhysRevLett.119.192502}. Otherwise, values from theoretical predictions \cite{Moeller.Pfeiffer.ea:2003} were used.

    %The astrophysical conditions used in the reaction network were chosen to match possible conditions of the recently observed
    %\cite{evans2017swift} 
    %blue kilonova following the GW170817 merger. 
    The thermodynamic evolution was parametrized assuming a free homologous expansion \cite{lippuner2015r}. Starting from an initial temperature of $6\,\mathrm{GK}$ and entropy of $10\,k_B/\mathrm{baryon}$ the expansion timescale was chosen to be $7\,\mathrm{ms}$. Qualitatively, our results are robust with respect to variations of
   the initial entropy and expansion timescale within a factor of two. 
	We calculate the abundance after $1\,\mathrm{Gyr}$ for a wide range of initial $Y_e$ between $0.28-0.43$, consistent with the lanthanide-free ejecta of the blue kilonova. 
	
In Fig.~\ref{fig:solar_r_compare_Ye} we show a subset of these
in comparison to the abundance (traditionally) assigned to the solar
\emph{r}-process \cite{Arnould2007}, obtained by subtraction of the
\emph{s}-processes from the solar abundance. The abundance pattern in
the region of the first \emph{r}-process peak is associated with large
uncertainties due to admixtures of weak- and main
\emph{s}-process. However for $A=79$ to $85$ species the
\emph{r}-process residuals have been estimated more precisely with
uncertainties of about $\approx 20$ to $50 \%$, see grey error band in
Fig.~\ref{fig:solar_r_compare_Ye}~and~\ref{fig:nucleo_error_bar}. This
is important for a precision study, because it allows for a fine
investigation and comparison of the production with BNS merger
calculations. We focus on this region, where we choose $^{82}$Se as a
reference isotope, because it is shielded from contributions of the
\emph{s}-process and requires a pure \emph{r}-process source. The outstanding features in this mass region are the abundance maxima at $A=80$ and $A=84$.

%Ref.~\cite{Hartmann1985} has studied nucleosynthesis in neutron-rich
%conditions similar to those explored in the present work. However,
%their calculations assumed nuclear statistical equilibrium (NSE). 
Nucleosynthesis under neutron-rich conditions, similar to those explored in the present work, has been studied assuming nuclear statistical equilibrium (NSE) \cite{Hartmann1985}.
In our full-scale network calculations we find that NSE provides a good
description of the abundances for temperatures above 4.5~GK. Below
this temperature the network and NSE abundance distributions result in
very different peak structures. As an example, for $Y_e=30/80=0.375$
NSE produces only $A=80$ nuclei while the network produces a broader
distribution of nuclei including peaks at $A=80$, $81$ and 84 (see
Fig.~\ref{fig:solar_r_compare_Ye}). In our network calculations $Y_e$
between $0.35-0.38$ provide the strongest contribution to the mass
region around the $A=80$ and $A=84$ abundance peaks. Lower $Y_e$
overproduce the $A=90-120$ region by more than one order of magnitude
and were therefore discarded. $Y_e$ above $0.39$ do not reach the
$A=84$ abundance maximum, as shown in
Fig.~\ref{fig:solar_r_compare_Ye}, or produce only reduced amounts of
$A=84$, as e.g. shown for $Y_e=0.41$, and were not considered further.

%These conditions are representative of the dynamical ejecta in a neutron star merger.

% * <stelios.nikas.ph@gmail.com> 2018-07-25T09:32:53.625Z:
    %which can be confirmed with the newly measured masses.
   %
   %Hydrodynamical simulations \cite{Wanajo.Sekiguchi.ea:2014} have also shown that these conditions can be found in a significant fraction of the ejected material from a BNS merger, if the effects of neutrino irradiation are take into account.  
   %   

In Fig.~\ref{fig:nucleo_error_bar} we compare our results with the
solar \emph{r}-process abundances in the region $A\approx 80-90$. We
include uncertainty bands showing the variation of the abundances
arising from the error bars of the masses of
$^{84,85}$Ga. Calculations within $Y_e= 0.35-0.38$ were combined with
equal weight.
  %prior to our measurement. 
The new $^{84,85}$Ga mass values affect the abundances of elements
with mass number $A = 82-87$ with the biggest impact on $A=83$, which
changes by about $\approx 15\%$ despite the small change in mass value
(see explanation of the formation in Sec.
\ref{sec:formation}). %This matches the expected high sensitivity of the \emph{r}-process to neutron-rich Ga isotopes as predicted by \cite{Brett2012, doi:10.1063/1.4867193}.
	
Furthermore, the uncertainty of the production of the \emph{r}-process-only reference isotope $^{82}$Se is significantly reduced to a level now comparable to the uncertainty of its \emph{r}-process residual, which is crucial for drawing quantitative conclusions about the production in this region. 
	  
For combinations of mass values leading to a low neutron separation
energy of
$^{85}$Ga, %the neutron capture flow is suppressed and more material remains below $A=84$. This way
the formation of a $A=84$ abundance peak is reduced (see lower limit
error band Fig.~\ref{fig:nucleo_error_bar}).  The new mass values
reduce the uncertainty of the final abundance sufficiently and the
formation of an abundance peak at $A=84$ becomes plausible.
% GMP: Unclear what this adds. The uncertanty band mentioned in the
% previous sentence includes the effects of everything including late
% time neutron captures. 
% In
% addition, the reduced mass uncertainty constrains late time neutron
% capture rates for the isotopes of Ga which are most abundant at the
% freeze-out, further constraining the production.
 
The calculations show in general a good agreement compared to the
solar \emph{r}-process abundance, particularly for the $A=80$ and
$A=84$ abundance peaks, but a strong overproduction at $A=81$,
possibly $A=86$ and $A=87$ and a reduced production at $A=90$.
 
  \subsection{Formation of the Final Abundance}
\label{sec:formation}

    \begin{figure}[tb]
\includegraphics[width=0.98\linewidth]{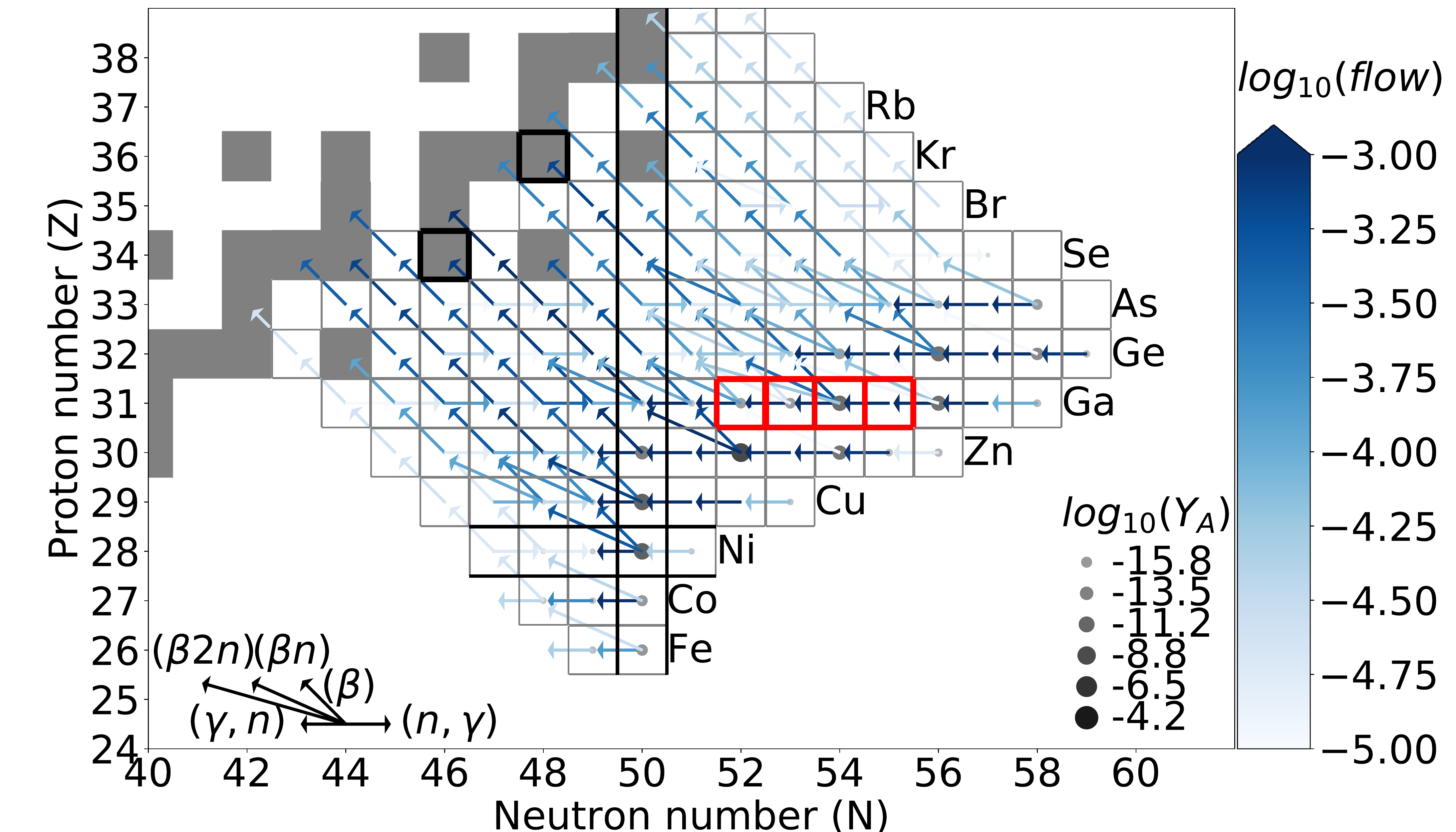}	
    \caption{%\todo{update to PRL standard font= Times New Roman} 
    (Color online) Time integrated reaction flows at $Y_e= 0.35$ from the freeze-out of the \emph{r}-process at a neutron-to-seed ratio of unity until the final abundances are established, relevant for the nucleosynthesis in the mass regions $A=78-86$. Most abundant nuclei at freeze-out are marked with solid circles. Grey filled squares indicate stable nuclei. Red squares indicate nuclei for which the reaction rates have been affected by the uncertainties of the mass value of $^{84}$Ga and $^{85}$Ga, while black ones indicate the abundance peaks at $A=80$ and $A=84$, corresponding to $^{80}$Se and $^{84}$Kr. Color coded arrows indicate the intensity of the flow.}
    \label{fig:flows}
   \end{figure}
  
The formation of the final abundance curve is illustrated by the
nuclear reaction flows shown in Fig.~\ref{fig:flows} based on the
calculation with $Y_e = 0.35$. They indicate the importance of
$\beta$-delayed neutron emission and late time neutron captures. When
taking the new mass values into account, most major nuclear physics
inputs required for the formation of the $A=84$ abundance peak in BNS
mergers are now in place. This is a unique situation at the first
\emph{r}-process peak and allows for the identification of remaining
key nuclei, whose masses and decay properties (half-lives and
$\beta$-delayed neutron emission) are urgently needed to fully
understand the possible formation of the first \emph{r}-process peak in BNS
merger calculations.

For the conditions considered here, the $A=80$ ($^{80}$Se) peak in the
solar \emph{r}-process residuals is mainly produced at the $N=50$
neutron magic number as $^{80}$Zn in the range
$Y_e \approx 0.36-0.37$. The $A=84$ ($^{84}$Kr) peak is produced for
the whole range of $Y_e$ values considered with contributions from the
neutron-rich Ga isotopes measured in the present work.  $^{84}$Ga,
having an odd-neutron number and a strong $\beta$-delayed neutron
emission branching of about $\approx 50 \%$ \cite{PhysRevC.95.054320},
does not contribute significantly to the final abundance of
$A=84$. The final abundance of $^{84}$Kr results mostly from the decay
of $^{85}$Ga. $^{85}$Ga is the most abundant species in the Ga
isotopic chain at the freeze-out of the \emph{r}-process as it has an
even number of neutrons. Due to its high $\beta$-delayed neutron
emission branching of $\approx 70 \%$ \cite{PhysRevC.97.054317} it
dominates the production of $^{84}$Kr.
	
The solar \emph{r}-process residuals for $A=86$ ($^{86}$Kr) and $A=87$
($^{87}$Rb) are very uncertain, hence the differences might arise from
uncertainties in the \emph{s}-process abundance
\cite{10.1093/mnras/sty1185}, which however can only account for some
of the discrepancy. To further investigate this overproduction precise
masses and $\beta$-delayed neutron emissions of $^{86,87}$Ga and
$^{86\text{--}88}$Ge are needed. We note that recently $\beta$-delayed neutron
emissions of $^{86,87}$Ga have been reported
\cite{PhysRevC.100.031302}, but are not yet included in our
calculations. Masses of more neutron-rich Ge will also confine
possible production of strontium in BNS mergers
\cite{watson2019identification}.

The $A=81$ ($^{81}$Br) abundance is produced mainly from
$\beta$-delayed neutron emission of $^{82}$Zn, whose half-life
exhibits some inconsistencies ($228(10)\,\text{ms}$
\cite{Madurga.Surman.ea:2012}, $178(2.5)\,\text{ms}$
\cite{Xu.Nishimura.ea:2014}, $155(20)\,\text{ms}$
\cite{Alshudifat.Grzywacz.ea:2016}), but more importantly, the masses
of $^{83,84}$Zn, that determine the neutron capture flow beyond
$^{82}$Zn and therefore its freeze-out abundance, are not
experimentally known and as such might alter the production of
$^{81}$Br.

\section{Conclusion}
\label{sec:conclusion}

In summary, using TITAN's MR-TOF-MS we were able to measure the mass of neutron-rich Ga isotopes $^{84}$Ga and $^{85}$Ga for the first time with %an average uncertainty of $\Delta m/m~\sim~4\times 10^{-7}$. 
uncertainties between 25--48~keV.
%Our precision mass measurements show the clear preservation of the $N=50$ \todo{the $N=50$ what?} within the Ga isotopic chain and the continuing of the expected smooth trend following a shell closure. %Therefore the AME2016 predictions relying on local trends for the mass of $^{84}$Ga and $^{85}$Ga have shown to be accurate. 
Performing \emph{r}-process nucleosynthesis calculations for
conditions possibly prevalent in the ejecta of the GW170817 BNS
merger, we show how light \emph{r}-process elements may be produced. In
our BNS merger calculations electron fractions with $Y_e=0.35-0.38$
contribute to the formation of the first \emph{r}-process abundance
peak. Under these conditions, we demonstrate that at moderate
neutron-rich conditions BNS merger calculations can produce the $A=84$
abundance feature of the solar system \emph{r}-process
residuals. Reducing nuclear physics uncertainties associated with
$^{84,85}$Ga isotopes is a step forward constrain nucleosynthesis of
light \emph{r}-process
elements. %, that keeps neutron star mergers in the running to explain the solar abundance.
In order to understand additional fine features of the abundance
pattern, e.g. the production of strontium in BNS merger, additional
nuclear physics uncertainties need to be addressed. In particular,
nuclear masses and decay properties of more neutron-rich Ge and Zn
isotopes are needed.

%we are able to cast more light on the production of first \emph{r}-process peak isotopes. Following the reaction flow from the freeze-out of the \emph{r}-process to stability, the formation of the $A=84$ abundance peak can be explained by the neutron star merger.  
%Depending on the astrophysical conditions of the \emph{r}-process, in particular in the case of higher neutron densities, masses and $\beta$-delayed neutron emission probabilities of more neutron-rich gallium and germanium isotopes are needed to understand the observed \emph{r}-process abundance pattern beyond the $A=84$ peak. 

% (this needs to move somewhere else, but nicely highlights the experimental challenge) Mass measurements of $^{85}$Ga were particular challenging due to the overwhelming amount of surface ionized $^{85}$Rb and were only possible due to the usage of the Ion-Guide Laser-Ion-Source (IG-LIS), suppressing the Rb contamination by $5-6$ orders of magnitude, and the Multiple-Reflection Time-of-Flight Mass-Spectrometer (MR-TOF-MS), capable of a almost $5$ orders of magnitude dynamic range. Thus the mass of $^{85}$Ga, produced at a rate of $\approx 1$~pps, could be determined out of a $\approx 10^{9}$~pps $^{85}$Rb background.      

\section*{Acknowledgement}
The authors want to thank the TRILIS group at TRIUMF for Ga beam development and beautiful operation of the IG-LIS ion source, J. Bergmann for the Massdata-Acquisition software package used for the MR-TOF-MS and discussions with F.-K. Thielemann are also acknowledged. This work was partially supported by Canadian agencies NSERC and CFI, U.S.A. NSF (grants PHY-1419765 and PHY-1614130) and DOE (grant DE-SC0017649), Brazil's CNPq (grant 249121/2013-1), United Kingdom's STFC (grants ST/L005816/1 and ST/L005743/1), the Canada-UK Foundation, German institutions DFG (grants FR 601/3-1 and SFB~1245 and through PRISMA Cluster of Excellence), BMBF (grants 05P15RGFN1 and 05P12RGFN8), the Helmholtz Association through NAVI (grant VH-VI-417), HMWK through the LOEWE Center HICforFAIR, by the JLU and GSI under the JLU-GSI strategic Helmholtz partnership agreement, the ChETEC COST Action (CA16117) supported by COST (European Cooperation in Science and Technology),  the Los Alamos National Laboratory and has been assigned report number LA-UR-18-28439. Los Alamos National Laboratory is operated by Los Alamos National Security, LLC, for the National Nuclear Security Administration of US Department of Energy (Contract DEAC52-06NA25396). TRIUMF receives federal funding via NRC-CNRC. 

\bibliographystyle{apsrev4-1}
\bibliography{library,jonas_refs}

\end{document}